\documentclass[twocolumn,preprintnumbers,secnumarabic,amsmath,amssymb,superscriptaddress,nofootinbib,floatfix]{revtex4}
\usepackage{varioref,exscale,latexsym,amsmath,amssymb}
\usepackage{graphicx}

\usepackage{graphicx}
\usepackage{dcolumn}
\usepackage{bm}

\def\beq{\begin{equation}}

\def\eeq{\end{equation}}

\def\beqa{\begin{eqnarray}}

\def\eeqa{\end{eqnarray}}

\begin{document}

\preprint{MPP-2009-24}

\title{{\bf Likely values of the Higgs vev}}

\medskip\
\author{John F. Donoghue}%
\email[Email: ]{donoghue@physics.umass.edu}
\affiliation{Department of Physics,
University of Massachusetts\\
Amherst, MA  01003, USA}
\author{Koushik Dutta}
\email[Email: ]{koushik@mppmu.mpg.de}
\affiliation{Max-Planck-Institut f\"ur Physik (Werner-Heisenberg-Institut)\\
F\"ohringer Ring 6,
D-80805, M\"unchen, Germany }
\author{Andreas Ross}
\email[Email: ]{andreas.ross@yale.edu}
\affiliation{Department of Physics,
Yale University \\
New Haven, CT 06520, USA }
\author{Max Tegmark}
\email[Email: ]{tegmark@mit.edu}
\affiliation{Department of Physics,
Massachusetts Institute of Technology  \\
Cambridge, MA 02139, USA }

\begin{abstract}
We make an estimate of the likelihood function for the Higgs vacuum
expectation value by imposing anthropic constraints on the existence
of atoms while allowing the other parameters of the Standard Model
to also be variable. We argue that the most important extra ingredients
are the Yukawa couplings, and for the
intrinsic distribution of Yukawa couplings we use the scale
invariant distribution which is favored phenomenologically. The
result is successful phenomenologically, favoring values close to
the observed vev. We also discuss modifications that can change
these conclusions. Our work supports the hypothesis that the
anthropic constraints could be the origin of the small value
of the Higgs vev.

\end{abstract}
\maketitle
\section{Introduction}

It is known that if the masses of the light quarks and the electron
were modestly different then nuclei and atoms would not exist \cite{damour, anthropic, agrawal}.
Because the masses of fermions are proportional to the Higgs vacuum
expectation value (vev), these bounds can be interpreted as
constraints on possible values of the Higgs vev if the other
parameters of the Standard Model are held fixed \cite{damour, agrawal}.
This observation is
interesting because it could provide an answer to one of the most
significant puzzles of the Standard Model - often called the fine
tuning problem or the hierarchy problem. If the underlying theory
allows the existence of different values of the Higgs vev, we would
only find ourself in a region of the universe that contains atoms,
and hence the vev may be constrained to a small range of possible
values. This provides a motivation for theories with multiple
possible values of the physical parameters, such as the string
theory landscape \cite{landscape}.

However, in theories in which the basic parameters can take on
multiple values, other parameters besides the Higgs vev will most likely
also be
variable. This would be the case in the string landscape picture.
Since the atomic constraints are really on the up quark, down quark
and electron masses, they translate to constraints on the product of
the fermion Yukawa couplings and the Higgs vev, not the vev
uniquely. Even without the exploration of specific theories,
we might hope that the rough conclusion is unchanged,
namely that the scale of the electroweak sector must be reasonably
close to the scale of the strong interactions in order that the
masses of the light quarks and electron - the product of the
electroweak interactions - be comparable to nuclear binding energies
- primarily due to strong interactions.

In this paper we consider the more general case of allowing other
parameters of the Standard Model to vary, and attempt to provide a
likelihood distribution for the Higgs vev. One might think that
this would be possible only with the knowledge of the full underlying theory,
but we will primarily use data for this purpose. As we will argue in
Sec. 2 and 3, this is
possible because it is the Yukawa couplings that
appear to have the most significant influence on the range of the vev. Because
there are many masses, we then have experimental indication of
the intrinsic probability distribution for the
Yukawa couplings. The observed
quark and lepton masses provide quite strong statistical evidence
that this distribution is close to scale invariant \cite{weight1, weight2}.
We will review this idea in Section 4 and
provide further evidence in its favor in Section 5.

Applying such a scale invariant probability distribution for the
Yukawa couplings, we investigate the likelihood distribution for
the Higgs vev. This result is obtained
by finding the relative probability that the $u,~d$ quarks and the electron (governed by the
scale invariant weight) fall in the anthropically allowed range.
In this case our result is developed and displayed in Sec. 6.
We can also study the effect of producing modest changes in our
underlying assumptions. These are studied in Sec 7. We present our conclusions in Sec 8.

We are aware of many limitations of our work. Besides the assumptions that we
state and explore, there are likely other effects (nucleosynthesis, cosmology, etc)
that come into play, especially once we consider significant changes in the
parameters of the Standard Model. However, one would expect that
possible further anthropic
constraints would only tighten the likelihood function in the
neighborhood of the physical value.  The goal of the present work is to obtain
a sense of whether or not the atomic constraints could be the origin of the
low value of the Higgs vev (within the context of landscape-like theories). Our work
can be viewed as an attempt to quantify this by looking at what may be the
dominant effects.
Overall our conclusion is that it remains plausible that the atomic
constraints are the origin of the low value of the Higgs vev.

\section{General framework}

The situation that we have in mind is similar to the string
landscape picture in which there are very many possible values of
each of the parameters. While in string theory the choices of
parameters are discrete, the results appear to be so densely packed
as to appear almost continuous\footnote{For example, it has been estimated
that there are $10^{100}$ string vacua reproducing the Standard
Model parameters within the present experimental error bars \cite{landscape}, and the
density of states would be equally high in the neighborhood of these
parameters.}. We then describe the ensemble of such states by an
intrinsic probability distribution or weight that specifies the
probability of finding different values of the parameters. These
probabilities would emerge from string theory and the weight encodes
the shape of the string landscape. Let us call this weight
$\rho(v,\Gamma_i,g_i)$ where $v$ is the Higgs vacuum expectation
value, $\Gamma_i$ are the Yukawa couplings, and $g_i$ stands for the
gauge couplings and all other parameters of the theory.

However, many combinations of the parameters do not lead to nuclei
and atoms. There is an intrinsic selection effect that we
would only find ourselves in
friendly regions that include atoms. The shape of the intrinsic probability
distribution in unfriendly regions of parameter space is then
completely irrelevant for us and we are only concerned with the
parameter subspace that leads to atoms. Let us denote by $A(
v,\Gamma_i, g_i)$ the function that is zero for all parameters
that do not lead to atoms and unity for those that do. We will refer
to this as the atomic function.
In principle, the atomic function could also take into account not
only the mere existence of atoms but also the probability of
a physical environment of sufficient complexity
developing with the atoms that are available with that
parameter set. As one moves around the parameter space, especially
near the allowed borders, greater or fewer numbers of atoms exist
and/or would be produced in the early universe. With a reduced or
enhanced set of atoms available, the resulting complexity
might be greatly reduced or enhanced. However, such considerations
are beyond our capabilities to calculate and we do not consider
them. Given the primary features uncovered below, it is unlikely
that modifying the boundaries of the atomic function would have a large effect
on our results. Moreover, it is certainly cleaner and more conservative to limit our
discussion to the general physical characteristics of atoms and nuclei.

With the atomic function we can obtain the total probability to
find atoms in the landscape
\begin{equation}
P(A) = \int dv ~ d\Gamma_i ~dg_i~ A(v,\Gamma_i,g_i) ~ \rho(v,\Gamma_i, g_i) \label{eq:1}
\end{equation}
where A denotes the existance of atoms. However, this total
probability is not a quantity of interest. A more interesting one is
constructed by omitting the integration over $v$,
\begin{equation}
L(v) = \int d\Gamma_i ~dg_i~ A(v,\Gamma_i,g_i) ~ \rho(v,\Gamma_i,
g_i) \label{eq:2}
\end{equation}
which we call the likelihood function. It is in fact the probability
density $\frac{dp}{dv}$ for atoms to exist.

Another useful assumption is the independence of parameters. This means
that we assume that the intrinsic probability function factorizes
into the product of separate weights. In formulas this implies
\begin{equation}
\rho(v,\Gamma_i, g_i) = \rho(v) \rho(\Gamma_e)\rho(\Gamma_u)
\rho(\Gamma_d)......
\end{equation}
This is at least partially motivated by the vastness of the string
landscape. If we hold all but one parameter fixed, there are likely
other allowed vacua with this last parameter scanning over its
allowed range. There is also some phenomenological evidence in favor
of this from the distribution of quark and lepton masses, which all
seem consistent with the same distribution. However, we note the approximate nature of
this feature in our discussion of Sec. 7.

There also could be an a-priori distribution for the Higgs vev, $\rho(v)$, which
is a property of the fundamental theory. We clearly do not know this function.
However, it is expected that the vev can take on values in a very large range, at
least up to a unification scale. Moreover, there are many additive contributions to $v$ that come from quantum corrections. These add linearly in the respective couplings and this suggests that the overall distribution could be Taylor expanded in $v$ about the observed value. If this is the case, then our considerations cover
only a very small portion of the allowed range, and we treat the a-priori distribution
as a constant in this narrow range. If the a-priori distribution in $v$ were to be highly
peaked in some direction, our results would be modified, and so this must count as an
uncertainty in our method. Therefore, our assumption will be $\rho(v) \sim$~constant, and unless we know more about the underlying theory,
we feel that this is the most reasonable assumption under which to proceed.

The quantity we will explore in detail below is the probability to find atoms for a given
value of $v$. It is obtained by taking a sample $v$ and drawing the other relevant parameters
randomly from the probability distributions we consider. Then we decide if the resulting configuration
can yield atoms or not. With a large sample for a fixed value of $v$ we can obtain the probability
of having atoms by dividing the number of times we obtained atoms by the total number of simulations.
Using the assumption of independence of the parameters introduced above, the quantity we obtain from
our simulations then is
\begin{equation}
 P(\text{A}|\text{given }v) = \! \int \! d\Gamma_i \int \! dg_j \! \left(\prod_{i,j} \rho(\Gamma_i) \rho(g_j) \! \right) \! A(\Gamma_i, g_j, v) \label{eq:4}.
\end{equation}
Now the likelihood function $L(v)$ of Eq. (\ref{eq:2}) is obtained by the product of
$P(\text{A}|\text{given }v)$ and the intrinsic probability distribution $\rho(v)$.
Under our assumption of a flat $\rho(v) \sim const.$ in the range of interest,
the likelihood function is then simply proportional to $P(\text{A}|\text{given }v)$
of Eq. (\ref{eq:4}).

From the shape of the likelihood function $L(v)$ we can infer which values of $v$ are
typical and which ones are highly improbable. Since $L(v)$ is a probability density
its shape itself is not a direct indicator of the most likely values of $v$. A peak
in $L(v)$ for example does not indicate the most likely values of $v$; more
meaningful quantities to give would be the median or other percentiles.
A simpler way to explore the order of magnitude of the most likely values of $v$
can be obtained by plotting our results for $L(v)$ on a log-log scale. Since any
probability distribution which has a finite value of its percentiles must fall of
faster than $1/v$ at large values of $v$, the log-log plots show us if and when the
likelihood function falls off faster than $1/v$. If present, this point is then a
reasonable estimate of the most likely values of $v$. If $L(v)$ does not fall off
faster than $1/v$, no constraints on the Higgs vev arise from the existance of atoms.

\section{Brief summary of atomic constraints}

To the extent that we understand how the Standard Model leads to the world that we observe, we should be able to
describe the world that would result if we instead used parameters different from, but in the neighborhood of,
those seen in Nature. Surprisingly, the structure of the elements changes dramatically for quite modest
changes in the quark masses. In a recent paper \cite{damour}, Damour and Donoghue have tightened and
summarized the anthropic constraints on
quark masses\footnote{See also \cite{anthropic}.}. Here we briefly summarize these results.

The first constraint which results from the binding of nuclei gives an upper bound on the sum $m_u+m_d$. The key feature here is
that the pion mass-squared is proportional to this sum of masses, and as the pion mass gets larger nuclear binding quickly becomes weaker.
The binding energy is small on the scale of QCD and is known to have
opposing effects from an intermediate range attraction and a shorter range repulsion. The attractive component,
heavily due to two pion exchange,
is the most sensitive to the pion mass and weakening it leads to a lack of binding of nuclei. From \cite{damour}
this constraint is
\begin{equation}
m_u +m_d \le 18 ~{\rm MeV}.
\end{equation}

The second constraint comes from the stability of protons. If protons could annihilate with electrons, $p + e^- \rightarrow n + \nu_e$,
hydrogen would not exist. The proton and neutron mass difference gets contributions from the
quark masses and from electromagnetic interactions. Using the best present estimates of these, the constraint becomes \cite{damour}
\begin{equation}
m_d-m_u - 1.67 m_e \ge 0.83 ~{\rm MeV}.
\end{equation}
The right hand side of the equation is linear in the electromagnetic fine structure constant. Modest variations in
this number would not influence our results significantly. In providing this constraint, it has been assumed that the
neutrino masses remain negligibly small. This feature is also anthropically required \cite{teg}.
\begin{figure}[t]
 \begin{center}
  \includegraphics[width=0.40\textwidth,height=!]{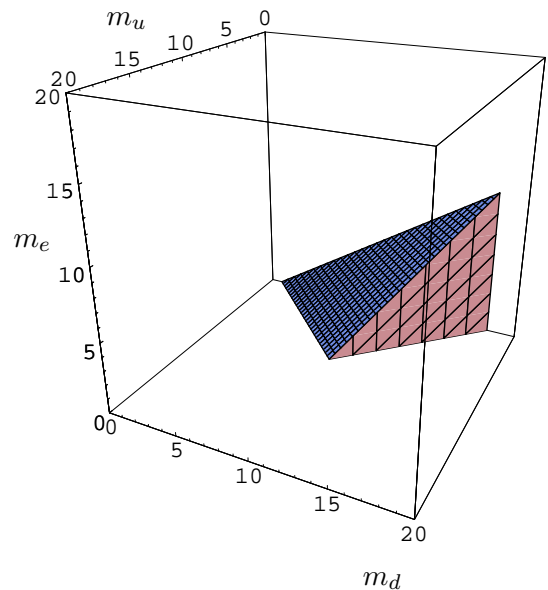}
 \end{center}
 \caption{\small{The anthropic constraints on $m_u, ~m_d,~m_e$ in MeV units.}}
 \label{3Dconstraints}
\end{figure}

These constraints are summarized in Fig.~\ref{3Dconstraints}. Note that the up quark and electron masses are able
to vary down to zero mass, while the down quark mass is constrained to be non-zero. The dimensional scale is set by the
$QCD$ scale $\Lambda_{QCD}$, so that these constraints could be rephrased in terms of dimensionless ratios $m_i/\Lambda_{QCD}$\footnote{In fact, $\Lambda_{QCD}$ serves as the comparison scale for all dimensional quantities in this work, so that the reference value for the Higgs vev $v_0$ is also
in units of $\Lambda_{QCD}$. }.

There are no known atomic constraints on the masses of the heavier quarks and leptons as long as they are
significantly heavier than the up, down and electron. Heavy quarks decouple from low energy physics
and have little influence on nuclei and atoms. Therefore, in our case the atomic function $A(v,\Gamma_i, g_i)$ reduces to $A(v, \Gamma_u, \Gamma_d, \Gamma_e)$. However, the Higgs vev does influence the mass of the $W$ gauge boson, and this
in turn influences the rates of weak processes. Since weak interactions play a significant role in the $pp$ cycle in stars, in particular
the crucial reaction $p+p \to d +e^+ +\nu$, raising the $W$ mass will slow the production of elements. So while atoms may exist, they may
not be produced in favorable numbers. This constraint is complicated also by the dependence on the cosmological parameters  governing stars.
In \cite{kribs}, a special scenario was constructed in which the W mass can be taken
to infinity if modifications are made to the usual picture of
nucleosynthesis\footnote{See, however, \cite{Clavelli} which raises problems with this scenario.}.
So while there may be additional constraints from the $W$ mass
which would further tighten the likelihood function, they have not been convincingly shown to be
tighter than those of the atomic function, so we will ignore them in our analysis. Because further constraints
lead to a narrowing of the likelihood function, omitting them is therefore a
more conservative approach.

\section{The distribution of quark and lepton masses}

In a landscape picture, the Yukawa couplings would not be uniquely determined
but would follow from some intrinsic distribution. Because there are enough masses,
the observed masses can give us insight into what this distribution is without
having to know the full underlying theory of the landscape \cite{weight1, weight2}\footnote{The Yukawa
interactions also influence quark mixing and the observed weight is consistent with the
hierarchy of weak mixing elements \cite{weight2, Gibbons}. There are also possible implications
for neutrino properties \cite{weight2, hall}.}.

The intrinsic probability distribution for a quark or lepton
Yukawa coupling is defined such that the fraction of values that appear at
coupling $\Gamma$ within a range $d \Gamma$ is $\rho(\Gamma)~d\Gamma$, with the normalization
\begin{equation} \label{eq:weightnorm}
1 = \int d\Gamma~\rho({\Gamma}).
\end{equation}
In particular, we have explored a set of power law weights \cite{weight1, weight2}
\begin{equation}
\rho(\Gamma) = \left\{
\begin{array}{cl}
\frac{N}{\Gamma^\delta}&\hbox{if }\Gamma_{\rm min}<\Gamma<\Gamma_{\rm max},\\
0&\hbox{otherwise},
\end{array}
\right.
\end{equation}
where the normalization constant is $N=(1-\delta)/[\Gamma_{\rm max}^{1-\delta} - \Gamma_{\rm min}^{1-\delta}]$ if $\delta\ne 1$ and
$N=1/\ln[\Gamma_{\rm max}/\Gamma_{\rm min}]$ if $\delta=1$.

Such simple power law weights require at least one endpoint (or two for $\delta = 1$) in order to be normalizable as in Eq. (\ref{eq:weightnorm}).
When we determine the endpoints for the Yukawa distributions at large and low values,
it is natural to use the renormalization group quasi-fixed point \cite{ross}  $\Gamma_{max}\sim 1.26$
as the upper limit. In \cite{weight1, weight2} a lower endpoint $\Gamma_{min} \sim 1.18 \times 10^{-6}$ which corresponds to $0.4 m_e$ was used. This is explored
more in the following section. With these ingredients, a likelihood analysis found that $\delta = 1.02 \pm 0.08$.

  \begin{figure}[t]
   \begin{center}
    \includegraphics[width=0.40\textwidth,height=!]{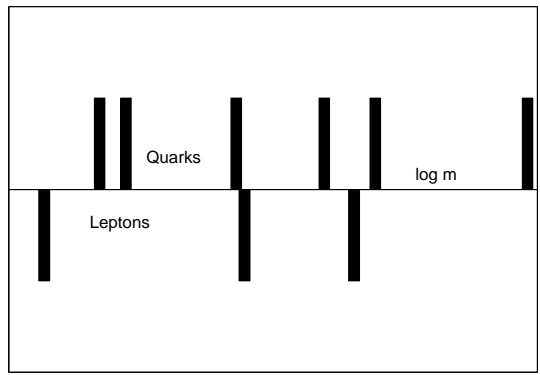}
   \end{center}
   \caption{\small{Quark and lepton masses, defined at the energy $\mu=M_W$, on a logarithmic scale. A
   scale invariant weight corresponds to a uniform distribution on this scale. We use this feature to
   describe the distribution of Yukawa couplings}}
   \label{scaleinvariant}
  \end{figure}

Of particular interest is the experimentally favored scale invariant distribution corresponds to a weight
with $\delta=1$:
\begin{equation}
\rho(\Gamma) = \frac{N}{\Gamma}.
\end{equation}
 While the evidence for the scale invariant weight
 is based on quantitative studies, the result can be seen
 qualitatively in Fig.~\ref{scaleinvariant}. A scale invariant distribution is one
 which is a random uniform population on a logarithmic scale.
 Fig.~\ref{scaleinvariant} shows the Yukawa couplings for the quarks and leptons, at
 the scale $\mu=M_W$, plotted on a logarithmic scale. The result
 appears visually to be consistent with this idea, and in practice a
 scale invariant weight is highly favored\footnote{Because the result is consistent
 with scale invariant, the choice of scale is not important and a similar
 result would be obtained using either the un-rescaled masses or with the Yukawa
 couplings defined at the Grand Unification scale.}. We will use this as our primary weight for our
analysis.

In exploring the uncertainties in the Higgs likelihood function, we will also
consider weights which have no lower bound on the Yukawa couplings. This is only possible
for $\delta<1$, if the distribution is to be normalizable. While the possibility $\Gamma_{min} =0$
is statistically disfavored (see next section), we
found in \cite{weight2} that the best fit value in that case is $\delta = 0.86_{-0.05}^{+0.04}$. We will use this in our tests of uncertainties in Sec 7.

\section{Further evidence on the fermion mass distribution} \label{sec:further}

\begin{figure}[t]
 \begin{center}
  \includegraphics[width=0.40\textwidth,height=!]{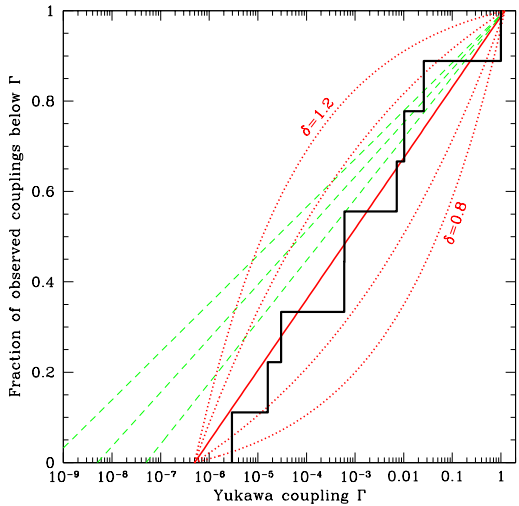}
 \end{center}
 \caption{\small{The staircase-like curve shows the empirical distribution function, defined as the fraction of the 9 observed Yukawa couplings that lie
 below a given $\Gamma$-value. The other curves show the cumulative distribution functions predicted by our power-law probability distribution Ansatz.
 All assume $\Gamma_{\rm max}=1.2$, and the straight lines correspond to the log-uniform $\delta=1$ case with different values of $\Gamma_{\rm min}$,
 the solid line having the best-fit value $\Gamma_{\rm min}=5\times 10^{-7}$.
 From top to bottom, the bent curves with the same endpoints take the $\delta$-values 1.2, 1.1, 1.0, 0.9 and 0.8, respectively.
 The $\delta=1$ line is seen to exchibit good consistency with the observed data.}}
 \label{cumulative}
\end{figure}

\begin{figure}[t]
 \begin{center}
  \includegraphics[width=0.40\textwidth,height=!]{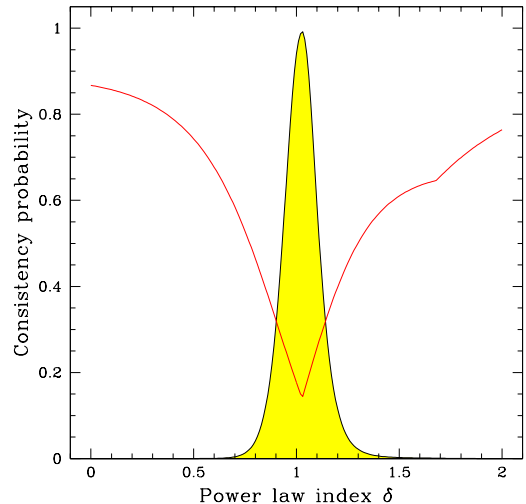}
 \end{center}
 \caption{\small{The probability that our model is consistent with 9 observed Yukawa couplings is shown as a function of
 the power law index $\delta$, assuming the best-fit value $\Gamma_{\rm min}=5\times 10^{-7}$}. The V-shaped curve shows the
 maximal difference between the predicted and observed distribution functions from Fig.~\ref{cumulative}.}
 \label{prob}
\end{figure}

\begin{figure}[th]
 \begin{center}
  \includegraphics[width=0.40\textwidth,height=!]{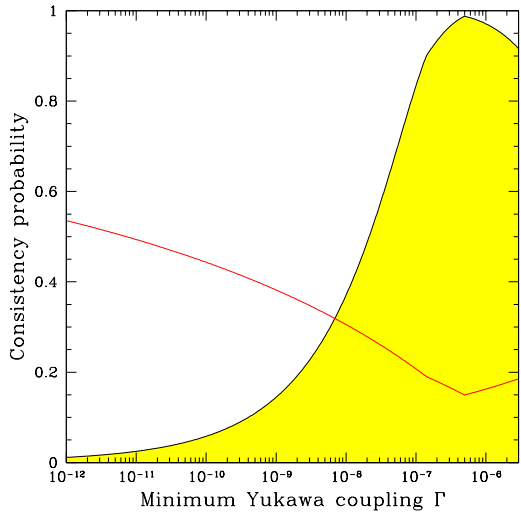}
 \end{center}
 \caption{\small{The probability that our model is consistent with 9 observed Yukawa couplings is shown as a function of
 the lower cutoff $\Gamma_{\rm min}$, assuming the scale-invariant power law index $\delta=1$. The V-shaped curve shows the
 maximal difference between the predicted and observed distribution functions from Fig.~\ref{cumulative}.}}
 \label{Gmin}
\end{figure}

Since the Bayesian result of Ref. \cite{weight2} $\delta = -1.02\pm 0.08 $ from a likelihood analysis does not address the question of whether
the best-fit distribution is actually consistent with the data, we have studied a set of frequentist Kolmogorov-Smirnov (KS) tests.
The observational input into a KS-test is the so-called empirical distribution function $F_{\rm obs}(\Gamma)$,
defined as the fraction of the 9 observed Yukawa couplings that lie
below a given $\Gamma$-value. This function is plotted in Fig.~\ref{cumulative}, and the horizontal location of the staircase-shaped steps correspond to
the 9 rank-ordered Yukawa couplings for ($e,~\mu,~\tau,~u,~d,~s,~c,~b,~t$), taken from the Particle Data Group compilation \cite{PDG}.
For any particular choice of our model parameters $(\delta,\Gamma_{\rm min},\Gamma_{\rm max})$, we can compute a predicted
cumulative probability distribution function
\begin{equation}
F(\Gamma)\equiv\int_{\Gamma_{\rm min}}^\Gamma \rho(\Gamma')d\Gamma'
\end{equation}
and compare how well it agrees with $F_{\rm obs}(\Gamma)$. Fig.~\ref{cumulative} illustrates this for variations in both
$\delta$ and $\Gamma_{\rm min}$.
The KS-test uses as a goodness-of-fit statistic the maximum vertical discrepancy $|F_{\rm obs}(\Gamma)-F(\Gamma)|$ between theory and observation.
The corresponding consistency probabability is shown in Figs.~\ref{prob} and~\ref{Gmin} for variations in $\delta$ and $\Gamma_{\rm min}$, respectively, keeping
a fixed upper cutoff $\Gamma_{\rm max}=1.2$.

The constraints on $\delta$ from Fig.~\ref{prob} nicely reproduce the previous likelihood result using frequentist methods.
Fig.~\ref{prob} shows that we can reject the null hypothesis that our model is correct at high significance if $\delta$ deviates substantially from unity,
and also that the data are perfectly consistent with $\delta=1$.
This approach also enables us to place a bound on the lower endpoint $\Gamma_{\rm min}$, as seen in Fig.~\ref{Gmin}.
Normalizability alone requires a lower endpoint when $\delta\ge 1$, but for any $\delta$-value, it is clear from Fig.~\ref{cumulative}
that the fit becomes poor if the left endpoint is dragged sufficiently far to the left of the leftmost data point.
Fig.~\ref{Gmin} shows that the best fit value is $\Gamma_{\rm min}\approx 5\times 10^{-7}$ for the scale-invariant case,
although the constraints are rather weak,
with $1-\sigma$ and $2-\sigma$ bounds (32\% and 0.045\% consistency probability) corresponding to
$\Gamma_{\rm min}\approx 7\times 10^{-9}$ and $\approx 5\times 10^{-11}$, respectively.

\begin{figure}[ht]
\begin{center}
\hspace{1cm}
  \begin{minipage}[t]{.06\textwidth}
    \vspace{0pt}
    \centering
    \vspace{78pt}
    \hspace*{0.4cm}
    \rotatebox{90}{$x_{min}$}
  \end{minipage}%
  \begin{minipage}[t]{0.6\textwidth}
    \vspace{0pt}
    \hspace{-4cm}
    \centering
    \includegraphics[width=0.60\textwidth,height=!]{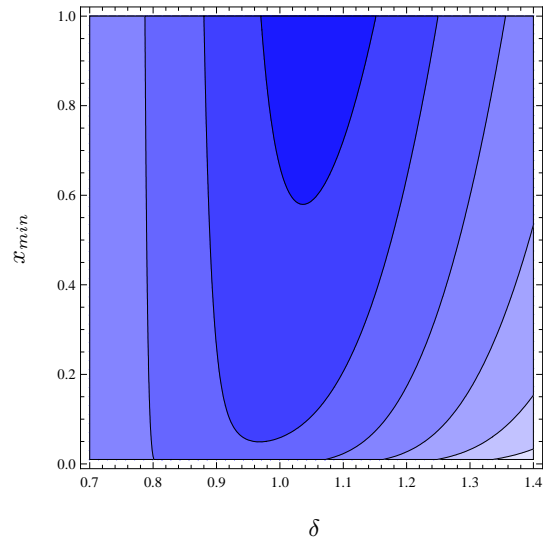}
  \end{minipage}
\end{center}
$\hspace*{21pt} \delta$
\caption{\small{Contour plot of the log-likelihood as a function of the lower endpoint for the Yukawa couplings
  $\Gamma_{min}$ and the exponent $\delta$. We normalize $\Gamma_{min}$ relative to the lowest observed Yukawa coupling, $\Gamma_e$,
  by defining $x_{min} = \Gamma_{min}/\Gamma_e$. The darkest area is the $1\sigma$ range, the second darkest area marks the $2\sigma$
  range etc. \vspace*{10pt}}}
 \label{mlow}
\end{figure}

In Ref. \cite{weight2} a Bayesian approach was employed where the dependence of the likelihood function
on the exponent $\delta$ was studied for a fixed $\Gamma_{min}$. Here we extend this analysis by investigating
also the dependence of the likelihood on $\Gamma_{min}$. As in \cite{weight2} we fix the upper endpoint
to be $\Gamma_{max} = 1.26$ which is motivated by the quasi-fixed point of the Standard Model. In Fig.~\ref{mlow}
we show the contour plot of the log-likelihood function as a function of the lower Yukawa endpoint
$\Gamma_{min}$ and the exponent $\delta$. The darkest area is the $1-\sigma$ range, the next one is the $2-\sigma$ range etc.
Here the $n-\sigma$ range is taken as the parameter space where the log-likelhood is at most $n^2/2$ below its maximum.

We see that for all values of $\delta$ the data favors a lower endpoint for the distribution,
and for any fixed value of $\delta$ the likelihood function increases monotonically up to the
highest possible value $\Gamma_{min} = \Gamma_e$.
The best fit, i.e. the highest likelihood, is found for $\delta = 1.06$ and $\Gamma_{min} = \Gamma_e$. Any
probability distribution with $\Gamma_{min} = \Gamma_e$ seems of course very unnatural since one out of the
nine measured Yukawas would lie exactly on the endpoint of the probability distribution, but the
likelihood analysis does not take that into account. For $\delta<1$ where the power law weights
do not require a lower endpoint $\Gamma_{min}$, the data says that such a scenario with
$\Gamma_{min} = 0$ is disfavored by over $2-\sigma$.

\section{The likelihood function for the Higgs vev}

We now combine our two key ingredients, the anthropic constraints and the
probability distribution for the Yukawa couplings which is phenomenologically
successful in describing the observed Yukawa couplings, in order to estimate
the likelihood distribution for the Higgs vev. Our approximation of the full problem,
under the assumptions described in Section 2, consists of
\begin{equation}
  L(v) = \int d\Gamma_i ~ A(v,\Gamma_u,\Gamma_d,\Gamma_e) ~ \rho(\Gamma_i)
\label{truncated}
\end{equation}
where a product over all charged fermions $i$  is understood. In comparison with Eq. (\ref{eq:2}), the gauge couplings have been dropped due to our focus
on the primary constraints due to the Yukawa couplings. The potential dependence on $v$ in $\rho(v, \Gamma_i...)$ is no longer present due to
the assumption that the intrinsic probability distribution in $v$ is roughly flat over our the allowed atomic window\footnote{we will return to these issues in our discussion of the uncertainties}. The atomic function $A(v, \Gamma_u,\Gamma_d,\Gamma_e)$
is summarized in Fig.~\ref{3Dconstraints}
and $\rho(\Gamma_i)$ is the probability distribution for the Yukawa couplings where we
use $\delta = 1$, $\Gamma_{min} = 0.4 \Gamma_e$ and $\Gamma_{max} = 1.26$.
The normalization of $L(v)$ is irrelevant, we are only interested in
estimating its shape. We calculate $L(v)$ numerically by randomly populating the Yukawa
couplings using the appropriate weight at different values of the vev. In particular
we generate a set of three Yukawa couplings for the up-type quarks, for the down-type quarks and for the leptons.
The smallest Yukawa coupling of each set is defined to be $\Gamma_u$, $\Gamma_d$ and $\Gamma_e$
respectively. The relative probability of satisfying the atomic constraints then yields the
likelihood function.

\begin{figure}[th]
 \begin{center}
  \includegraphics[width=0.40\textwidth,height=!]{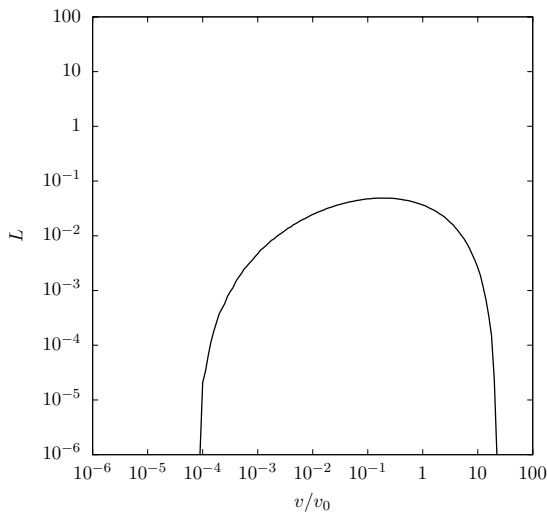}
 \end{center}
 \caption{\small{The likelihood function $L$ for the Higgs vev as a function of
 $v/v_0$ ($v_0$ is the observed Higgs vev of 246 GeV), constructed with our favored scale invariant weight. The normalization
 gives the fraction of simulations which satisfied the anthropic constraints.}}\label{ude}
\end{figure}

\begin{figure}[th]
 \begin{center}
  \includegraphics[width=0.40\textwidth,height=!]{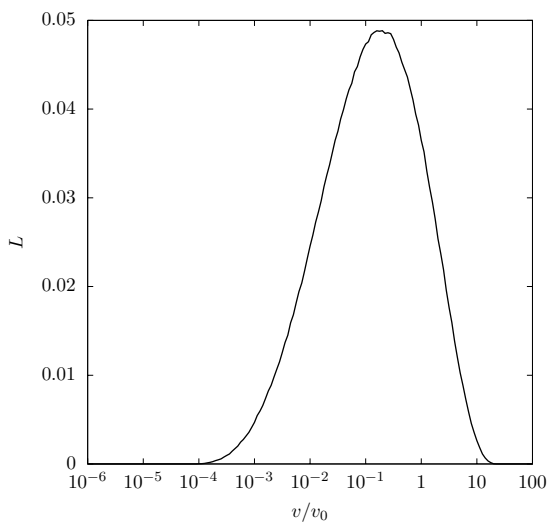}
 \end{center}
 \caption{\small{This shows the same likelihood function $L$ as in Fig. \ref{ude}, but with the vertical axis on a linear scale. }}\label{ude2}
\end{figure}

\begin{figure}[th]
 \begin{center}
  \includegraphics[width=0.40\textwidth,height=!]{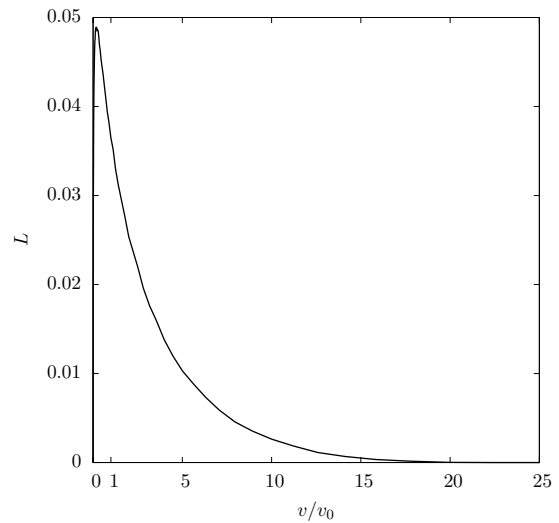}
 \end{center}
 \caption{\small{The likelihood function $L$ shown in Fig.\ref{ude} and Fig. \ref{ude2} is shown here with both the horizontal and vertical axes on a linear scale.}}\label{ude3}
\end{figure}

Let us briefly reiterate here the main assumptions which go into our analysis and explain the logic of how
the application of a scale invariant probability distribution for the Yukawa couplings can yield a constraint on
the scale of the Higgs vev. First of all, we note that we extract the observed Higgs vev
from measurements other than the fermion masses, such as $M_W$. Now, our crucial assumption
that the Higgs vev and the Yukawa couplings are statistically independent is used to infer
the probability distribution for the Yukawa couplings independently of the Higgs vev.
In practice, it essentially means that we infer the scale invariant weight for the Yukawa couplings directly
from the observed scale invariant weight for the quark and lepton masses for a fixed value of the Higgs vev.
The probability distribution for the Yukawa couplings is taken to be a power law with an exponential
close to the scale invariant case of $\delta = 1$. For $\delta = 1$ it requires lower and upper
endpoints.
Whereas the lower endpoint of the distribution is inferred from the measured Yukawa couplings, for the upper endpoint the renormalization group quasi-fixed point has been used.
Again, the assumption of statistical
independence is crucial when we extract these endpoints and use them universally for any
Higgs vev. Thus, we know that the Yukawa couplings have to not only be uniformly distributed on a log scale
in a stretch extending over roughly six orders of magnitude, but we know where on the log scale this stretch is located.
Finally, a scale enters through the atomic constraints on the light fermion masses, and together with the
probability distribution of Yukawa couplings, it yields our constraints on the Higgs vev.
Consider for example a large Higgs vev of $10^{16}~{\rm GeV}$. Since our Yukawa couplings from a scale invariant weight
are required to lie roughly between $10^{-6}$ and $1$, it would be impossible to get light fermion masses
in the ${\rm MeV}$ range for such a large value of the Higgs vev.

For our main result, we consider the case seen in Nature where only the $u,~ d$ quarks and the electron fall within the
anthropic window - all others quarks and leptons are heavier and should not be part of stable atoms. We discuss alternatives in the next section. We implement this constraint by requiring that the second lightest up-type quark, the $c$ quark, does not lie within the anthropic window
sketched in Fig.~\ref{3Dconstraints} when the $m_u$ axis is replaced by $m_c$, and analogously for the second lightest down-type quark.

For our favored scale invariant weight, the result is shown as Fig.~\ref{ude}, \ref{ude2} and \ref{ude3} using log-log, log-linear and linear-linear coordinate axes.
This is our primary result. We see that the distribution is peaked near the value $v_0$ observed in
Nature and it extends over several orders of magnitude. The median value in this distribution is $v = 2.25~ v_0$. The $ 2-\sigma$ range extends from $0.10~ v_0$ to $11.7~ v_0$.
We observe that there is a steep upper cutoff in the allowed values of the vev which comes
from the lower endpoint $\Gamma_{min}$ present in the scale invariant distribution of the
Yukawa couplings. The relevant point where the likelihood
function falls off faster than $1/v$ is located at a few times $v_0$. We
conclude that $v_0$ would be a very typical value for the Higgs vev
whereas values $\gg 10 \times v_0$ would be very unlikely.

\section{Uncertainties}

\begin{figure}[t]
   \begin{center}
    \includegraphics[width=0.40\textwidth,height=!]{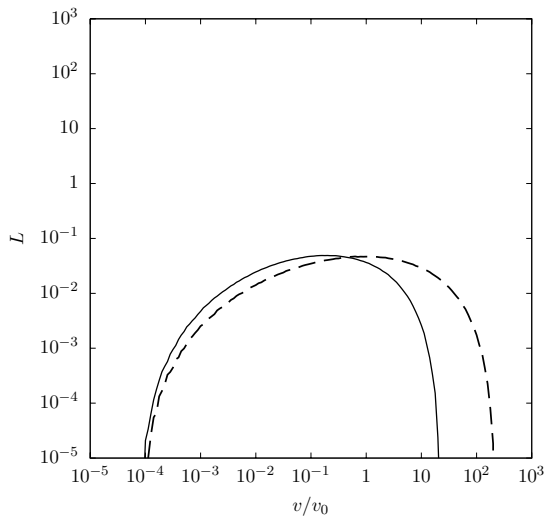}
   \end{center}
   \caption{\small{The likelihood function for different values of the lower endpoint in the Yukawa distribution,
                   $\Gamma_{min} = 0.4 \Gamma_e$ (solid) and $\Gamma_{min} = 0.04 \Gamma_e$ (dashed).}}
   \label{diffmlow}
\end{figure}

In this section, we consider the changes in the likelihood function if we modify some of the features of our analysis. The two greatest effects come from the variation or removal of the lower endpoint in the fermion mass distribution, and from the possibility of extra quarks or leptons within the anthropic window. The first of these has the potential to significantly modify our results.

First we can consider the changes if we use a different value of the lower cutoff. Changing the lower endpoint
by one order of magnitude within the context of the scale invariant weight produces the modification shown
in Fig.~\ref{diffmlow}. The plot shows that for $\Gamma_{min} = 0.04 \Gamma_e$ the likelihood distribution
favors larger values of the Higgs vev. In this case, smaller allowed values of the Yukawa couplings
are compensated by larger values of the Higgs vev to satisfy the atomic constraint. We see that the
qualitative features of the spectrum remain. However, the value of $v$ where the likelihood starts to falls off faster
than $1/v$ is roughly 10 times higher if we divide $\Gamma_{min}$ by a factor of 10. We clearly need to address this issue
of how much the most likely Higgs vev depend on the cutoff at low Yukawa couplings in the weight.

One major concern about the result of the previous section is that the shape of the likelihood function
is determined at large values of the vev by the fact that the scale invariant weight has a lower endpoint
to the Yukawa distribution. We address this issue by considering a power weight with $\delta$ less than unity, with no lower endpoint.
Thus the power law behavior of the weight at low Yukawa couplings is not cut off at all and extends all
the way down to zero. The Yukawa couplings can therefore become arbitrarily small, without being in violent
disagreement with the overall fermion mass distribution.
Such a distribution has the potential to allow arbitrarily large values for the vev. In these cases, there would be
situations where a high value of the vev is counterbalanced by very small Yukawas in order to satisfy
the atomic constraints.

\begin{figure}[t]
 \begin{center}
  \includegraphics[width=0.40\textwidth,height=!]{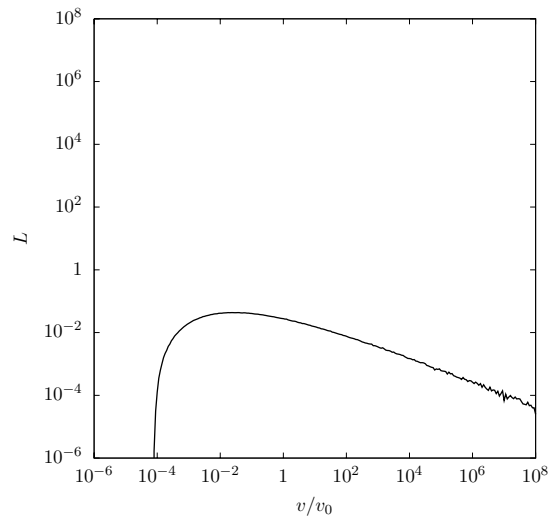}
 \end{center}
 \caption{\small{The likelihood function resulting from a power law weight of exponent $\delta=0.86$ without any lower endpoint $\Gamma_{min}$.}}
 \label{delta86}
\end{figure}

In Fig.~\ref{delta86}, we show the
likelihood function that is obtained for $\delta =0.86$ and $\Gamma_{min} = 0$.
While the maximum of the likelihood function remains close to the observed
value we see that it never falls off faster than $1/v$ in the region studied which extends up to $10^8 \times v_0$.
That means the most likely values of the Higgs vev in this scenario would be very large. Because the power law weight is
valid down to zero Yukawa couplings, we lose the constraints for the Higgs vev to be in the neighborhood of the observed $v_0$.
While our analysis in Sec. \ref{sec:further} has shown that $\Gamma_{min} = 0$ is disfavored by over
$2\sigma$ this issue remains a serious caveat. It motivates further top-down studies of properties of
the string theory landscape in order to identify the existence of a lower cutoff in the quark mass weight.

\begin{figure}[t]
 \begin{center}
   \includegraphics[width=0.40\textwidth,height=!]{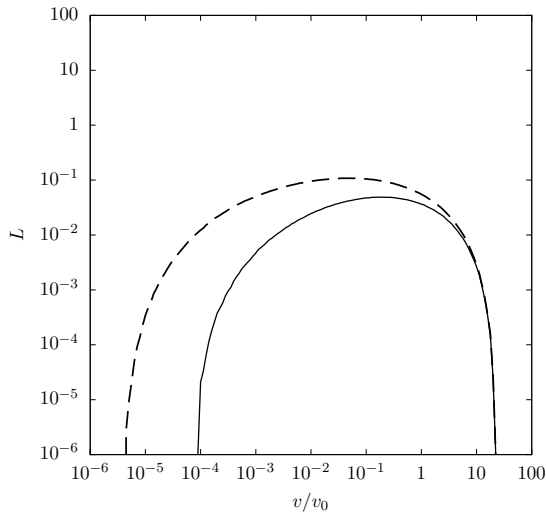}
\end{center}
\caption{\small{The likelihood function allowing any number of light quarks in the anthropic window (dashed),
along with that allowing $u$, $d$, $e$ only (solid).}}
 \label{both}
\end{figure}

Another uncertainty is the interesting possibility that more quarks beyond $u,~d,~ e$ fall in the atomic window.
If we treat all the quarks independently as we have been doing we find that it is
reasonably common that more quarks do fall in the allowed atomic window.
For example, with the scale invariant weight the likelihood function including extra quarks in this range is
shown in Fig.~\ref{both}. The allowance of extra small masses to fall in the atomic window makes the distribution
peaked around smaller values in comparison to when we do not allow them to fall inside the window. The value of $v$ where
the likelihood starts falling off faster than $1/v$ and therefore the most likely Higgs vev remains almost unchanged.
We do not see the disadvantage of this situation for the existence of atoms.

\begin{figure}[t]
 \begin{center}
   \includegraphics[width=0.40\textwidth,height=!]{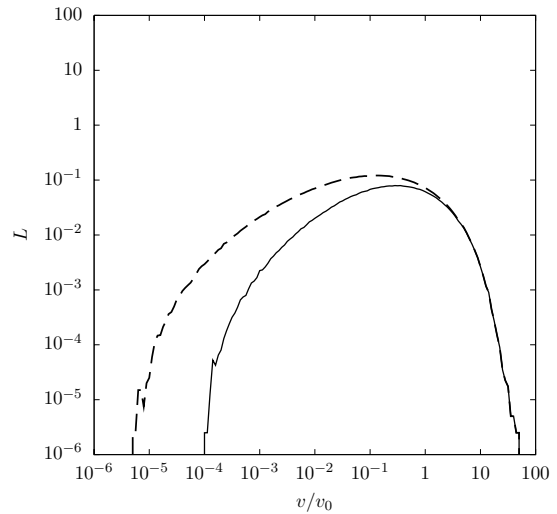}
\end{center}
\caption{\small{The likelihood function resulting from a biunitary diagonalization of the Yukawa matrices
                allowing any number of light quarks in the anthropic window (dashed),
                along with that allowing $u$, $d$, $e$ only (solid).}}
 \label{likematrix}
\end{figure}

However at this point it is useful to recognize a known flaw in our approximation that
the Yukawa distributions of each flavor are treated as independent. Even if the original
Yukawa couplings of the theory are distributed independently, the final output governing masses
will not be independent. The original Yukawas exist in a $3\times 3$ complex matrix for each charge sector. The
diagonalization of this matrix yields the final eigenvalues as well as rotation angles that go into
the CKM and PMNS weak mixing matrices. It is well known that upon diagonalization, the
eigenvalues of a matrix repel each other. In random matrix theory this leads to a repulsion of the
final eigenvalues. For our case, this says that the $u,~d,~e$ distributions will be independent,
since they come from different charge sectors, hence different matrices. However, the likelihood
of two quarks {\it of the same charge} falling in the allowed atomic range will be decreased by
this repulsion. We have studied this effect by generating random Yukawa couplings in a $3\times 3$
matrix and diagonalizing. As discussed in \cite{weight2}, in this case in order to
approximate a scale invariant distribution of the final eigenvalues we start with an initial weight with
$\delta=1.16$, and we use $\Gamma_{min} = 0.4 \Gamma_e$.
The resulting likelihood function is shown in Fig.~\ref{likematrix}, where the solid curve only allows
$u$, $d$, $e$ in the anthropic window and where the dashed curve results from allowing any number of
light quarks in the anthropic window. While the falloff at higher $v$ is now less steep than in the
result from diagonal simulations without matrix diagonalizations of Fig.~\ref{both}, it is clearly
falling off much faster than $1/v$ and the most likely value of $v$ would be below $10 \times v_0$.
Comparing the result from the diagonalization of the Yukawa matrices in Fig.~\ref{likematrix} with the
corresponding result from diagonal simulations without matrix diagonalizations shown in Fig.~\ref{both},
we note that the two curves in Fig.~\ref{likematrix} are closer together than the ones in
Fig.~\ref{both} which is due to the repulsion of eigenvalues in the matrix diagonalization case.

Since our work is based on statistical independence
of $v$ and $\Gamma_i$, i.e. $\rho(v, \Gamma_i) = \rho(v) \rho(\Gamma_i)$, there
could be even more dramatic problems if there were correlations between
$v$ and $\Gamma_i$. For example, if in our weight $\rho(\Gamma_i)$ both endpoints $\Gamma_{min}, \Gamma_{max}$ were proportional to $1/v$,
there would be no anthropic constraints on $v$ since varying $v$ would keep the distribution of the masses the
same. Because we measure the Yukawa couplings at a single value of $v$, we cannot address this.

We have mentioned previously the possibility that further anthropic constraints could come into play. These could change the detailed shape of the likelyhood function because they could eliminate regions of parameter space which are somewhat different from our world. However, since our parameters are clearly consistent with the constraints, the elimination of other regions would likely narrow the resulting likelihood function for $v$.
This would change the shape of the function, but would not modify our basic conclusions about the compatibility of our value of $v$ with the allowed anthropic range.

One might also wonder if the likelihood function should also take into account the other
great anthropic constraint, that on the clumping of matter in the
universe to form stars and planets, which limits in particular the
cosmological constant. The likelihood function for the cosmological
constant has been studied by Vilenkin and Garriga \cite{garriga}. The parameters of
the Standard Model do of course also influence the cosmological
constant. For example a shift in the value of the up quark mass (or the Higgs vev) by one
part in $10^{40}$ would shift the cosmological constant by 100\%.
However, for the cosmological constant to have an anthropic
selection, the possible values of $\Lambda$ must be densely packed
and there should be other slightly different combinations of
parameters that yield an anthropically allowed cosmological
constant. For the rather narrow range of $v$ that we probe, it seems very
reasonable
that the clumping of matter constraint has little influence on the
likelihood of the Higgs vev.

\section{Conclusions}
The fundamental question that we are addressing is whether it is plausible that
it is the atomic constraints that explain the low value of the Higgs vev in
landscape theories. This is known to be the case at fixed values of the Yukawa couplings, but
since these parameters also may vary in landscape theories, and since their magnitudes seem to be
peaked at low values, the answer is less obvious in a more general context. We have used experimental
information on the distribution of masses to address this issue. We find that even if the
Yukawa couplings are allowed to scan in a way that is favored phenomenologically, the likely values of the Higgs vev are close to
the one observed in Nature. This supports the hypothesis that these constraints favor
Higgs values similar to ours.

This can be interpreted as a motivation for further exploration of landscape theories. The
value of the Higgs vev and that of the cosmological constant are two great ``fine-tuning'' problems
of the fundamental interactions, and the presence of a landscape
would change the way that we approach the issue of fine-tuning \cite{Donoghue:2007zz}.
This is because both of these problems appear to have plausible resolutions through anthropic
constraints that are appropriate for landscape theories\footnote{However,
the ``strong CP problem'' is a fine-tuning puzzle
that does not appear to have an anthropic resolution \cite{strongCP}.}.

In \cite{kribs}, it was argued that the anthropic constraints on the quark masses
cannot be used to constrain the Higgs vev, by constructing a plausible scenario in which
the weak interactions do not appear. In the context of our exploration in this paper,
this could be realized by taking $v \to \infty,~ \Gamma \to 0$, with their product fixed.
This situation may arise in our scenario for power law weights with exponent $\delta < 1$
which extend down to zero Yukawa couplings. Such weights are disfavored by more than $2-\sigma$
in comparison to weights with a lower cutoff $\Gamma_{min}$ of the order of $\Gamma_e$
as we infered from the measured fermion spectrum. Nevertheless they are a serious caveat to
an anthropic constraint for the Higgs vev. For the prefered weights with a lower cutoff
$\Gamma_{min}$ of the order of $\Gamma_e$ the most likely values are close to that seen in Nature.

One of the strengths of our approach is that it does not rely on the ultraviolet completion of the
fundamental underlying theory. Aside from our assumption of statistical independence, this input
comes from the data on the quark and lepton masses. Both the issues of statistical independence and
of weights without a lower cutoff could possibly be addressed in top-down studies of the landscape.

The likelihood function that we have constructed is an estimator that tries to quantify
the effect of the possible variation of the fundamental parameters on the range of allowed
values for the Higgs vev. Further understanding of anthropic constraints may be able to
narrow the likelihood function further. Even if the range is
narrowed, this is more of a consistency check than a prediction
of landscape theories, since we already know that the observed value of the vev is
anthropically allowed. However, it does provide further motivation for landscape theories and
suggests that within these theories the hierarchy and fine-tuning problems associated
with the vev are not as serious obstacles as they are in other theories.

\section*{Acknowledgments}
J.D. acknowledges support from the National Science Foundation grant PHY- 055304.
K. D would like to acknowledge the financial support by the DFG cluster of excellence
``Origin and Structure of the Universe''.
A.R has been supported in part by the Department of Energy grant No. DE-FG-02-92ER40704.
MT was supported by NSF grants AST-0134999 and AST-05-06556,
a grant from the John Templeton foundation and a fellowships from the David and Lucille Packard Foundation. In addition both J.D. and M.T. thank the
Foundational Questions Institute for support and for conference travel that aided this collaboration.


\begin{thebibliography}{99}

\bibitem{damour}
  T.~Damour and J.~F.~Donoghue,
  ``Constraints on the variability of quark masses from nuclear binding,''
  Phys.\ Rev.\  D {\bf 78}, 014014 (2008)
  [arXiv:0712.2968 [hep-ph]].

\bibitem{anthropic} J. Barrow and F. Tipler, {\it The Anthropic
Cosmological Principle} (Clarendon Press, Oxford, 1986). \\
C.~J.~Hogan,
  ``Why the universe is just so,''
  Rev.\ Mod.\ Phys.\  {\bf 72}, 1149 (2000)
  [arXiv:astro-ph/9909295]. \\
R.~N.~Cahn, ``The eighteen arbitrary parameters of the standard
model in your everyday life,'' Rev.\ Mod.\ Phys.\ {\bf 68}, 951
(1996).\\
L.~B. Okun, Usp. Phys. Nauk, {\bf 161}, 177 (1991)(Soviet Phys. Usp.
{\bf 34}, 818).   \\
  R.~L.~Jaffe, A.~Jenkins and I.~Kimchi,
  ``Quark Masses: An Environmental Impact Statement,''
  Phys.\ Rev.\  D {\bf 79}, 065014 (2009)
  [arXiv:0809.1647 [hep-ph]].




\bibitem{agrawal}
V.~Agrawal, S.M.~Barr, J.F.~Donoghue and D.~Seckel, ``Anthropic
considerations in multiple-domain theories and the scale of
electroweak symmetry breaking,'' Phys.\ Rev.\ Lett.\ {\bf 80}, 1822
(1998)
hep-ph/9801253. \\
V.~Agrawal, S.M.~Barr, J.F.~Donoghue and D.~Seckel, ``The anthropic
principle and the mass scale of the standard model,'' Phys.\ Rev.\
{\bf D57}, 5480 (1998) hep-ph/9707380.

\bibitem{landscape}
M.~R.~Douglas, ``The statistics of string / M theory vacua,'' JHEP
{\bf 0305}, 046
(2003) [arXiv:hep-th/0303194].\\
S.~Ashok and M.~R.~Douglas, ``Counting flux vacua,'' arXiv:hep-th/0307049.\\
L.~Susskind, ``The anthropic landscape of string theory,'' arXiv:hep-th/0302219.\\
R.~Kallosh and A.~Linde, ``M-theory, cosmological constant and
anthropic principle,'' Phys.\ Rev.\ D {\bf 67}, 023510 (2003)
[arXiv:hep-th/0208157].\\
R.~Bousso and J.~Polchinski, ``Quantization of four-form fluxes and
dynamical neutralization of the  cosmological constant,'' JHEP {\bf
0006}, 006 (2000) [arXiv:hep-th/0004134].

\bibitem{weight1}
  J.~F.~Donoghue,
  ``The weight for random quark masses,''
  Phys.\ Rev.\  D {\bf 57}, 5499 (1998)
  [arXiv:hep-ph/9712333].

\bibitem{weight2}
  J.~F.~Donoghue, K.~Dutta and A.~Ross,
  ``Quark and lepton masses and mixing in the landscape,''
  Phys.\ Rev.\  D {\bf 73}, 113002 (2006)
  [arXiv:hep-ph/0511219].

\bibitem{teg}
  M.~Tegmark, A.~Vilenkin and L.~Pogosian,
  ``Anthropic predictions for neutrino masses,''
  Phys.\ Rev.\ D {\bf 71}, 103523 (2005)
  [arXiv:astro-ph/0304536].








\bibitem{kribs}
  R.~Harnik, G.~D.~Kribs and G.~Perez,
  ``A universe without weak interactions,''
  Phys.\ Rev.\  D {\bf 74}, 035006 (2006)
  [arXiv:hep-ph/0604027].

  \bibitem{Clavelli}
  L.~Clavelli and R.~E.~.~White,
  ``Problems in a weakless universe,''
  arXiv:hep-ph/0609050.





\bibitem{Gibbons}

  G.~W.~Gibbons, S.~Gielen, C.~N.~Pope and N.~Turok,
  ``Measures on Mixing Angles,''
  Phys.\ Rev.\  D {\bf 79}, 013009 (2009)
  [arXiv:0810.4813 [hep-ph]].
 G.~W.~Gibbons, S.~Gielen, C.~N.~Pope and N.~Turok,
  ``Naturalness of CP Violation in the Standard Model,''
  Phys.\ Rev.\ Lett.\  {\bf 102}, 121802 (2009)
  [arXiv:0810.4368 [hep-ph]].

\bibitem{hall}
 L.~J.~Hall, M.~P.~Salem and T.~Watari,
  ``Statistical Understanding of Quark and Lepton Masses in Gaussian
  Landscapes,''
  Phys.\ Rev.\  D {\bf 76}, 093001 (2007)
  [arXiv:0707.3446 [hep-ph]].  \\
 L.~J.~Hall, M.~P.~Salem and T.~Watari,
  ``Quark and Lepton Masses from Gaussian Landscapes,''
  Phys.\ Rev.\ Lett.\  {\bf 100}, 141801 (2008)
  [arXiv:0707.3444 [hep-ph]]. \\
 L.~J.~Hall, H.~Murayama and N.~Weiner,
  ``Neutrino mass anarchy,''
  Phys.\ Rev.\ Lett.\  {\bf 84}, 2572 (2000)
  [arXiv:hep-ph/9911341].\\
 N.~Haba and H.~Murayama,
  ``Anarchy and hierarchy,''
  Phys.\ Rev.\ D {\bf 63}, 053010 (2001)
  [arXiv:hep-ph/0009174].\\
 A.~de Gouvea and H.~Murayama,
  ``Statistical test of anarchy,''
  Phys.\ Lett.\ B {\bf 573}, 94 (2003)
  [arXiv:hep-ph/0301050].\\
 J.~R.~Espinosa,
  ``Anarchy in the neutrino sector?,''
  arXiv:hep-ph/0306019.\\
 L.~J.~Hall, M.~P.~Salem and T.~Watari,
  ``Neutrino mixing and mass hierarchy in Gaussian landscapes,''
  Phys.\ Rev.\  D {\bf 79}, 025010 (2009)
  [arXiv:0810.2561 [hep-th]].




\bibitem{ross}
B.~Pendleton and G.~G.~Ross,
  ``Mass And Mixing Angle Predictions From Infrared Fixed Points,''
  Phys.\ Lett.\ B {\bf 98}, 291 (1981). \\
 C.~T.~Hill,
  ``Quark And Lepton Masses From Renormalization Group Fixed Points,''
  Phys.\ Rev.\ D {\bf 24}, 691 (1981).\\
C.~T.~Hill, C.~N.~Leung and S.~Rao,
  ``Renormalization Group Fixed Points And The Higgs Boson Spectrum,''
  Nucl.\ Phys.\ B {\bf 262}, 517 (1985).\\
 M.~Lanzagorta and G.~G.~Ross,
  ``Infrared fixed points revisited,''
  Phys.\ Lett.\ B {\bf 349}, 319 (1995)
  [arXiv:hep-ph/9501394].






\bibitem{PDG}
  S.~Eidelman {\it et al.}  [Particle Data Group],
  ``Review of particle physics,''
  Phys.\ Lett.\ B {\bf 592}, 1 (2004).





\bibitem{garriga}
    J.~Garriga and A.~Vilenkin,
    ``Solutions to the cosmological constant problems,''
    Phys.\ Rev.\  D {\bf 64}, 023517 (2001)
    [arXiv:hep-th/0011262].

%
\bibitem{Donoghue:2007zz}
  J.~F.~Donoghue,
  ``The fine-tuning problems of particle physics and anthropic mechanisms,''
  arXiv:0710.4080 [hep-ph].
  'Universe or Multiverse' ed. by Bernard Carr pp 231-246 (Cambridge Univ. Press).

\bibitem{strongCP}
 J.~F.~Donoghue, ``Dynamics of M theory vacua'', Phys.\ Rev.\ D
 {\bf 69}, 106012 (2004) [Erratum-ibid.\ D {\bf 69}, 129901 (2004)]
 arXiv:hep-th/0310203].  \\
 T.~Banks, M.~Dine and E.~Gorbatov,
``Is There A String Theory Landscape,''
 JHEP {\bf 0408}, 058 (2004).
 [arXiv:hep-th/0309170].
 \\
L.~Ubaldi,``Effects of theta on the deuteron binding energy and the triple-alpha
process,''   Phys.\ Rev.\  D {\bf 81}, 025011 (2010).
 arXiv:0811.1599 [hep-ph].

\end{thebibliography}
\end{document}